\documentclass[bibyear]{aa} 
\usepackage{natbib}
\usepackage{multicol}
\usepackage{longtable}
\usepackage{caption}
\usepackage{graphicx}
\usepackage{subcaption}
\renewcommand\thesubfigure{\arabic{subfigure}}
\usepackage{txfonts}
\usepackage{hyperref}
\begin{document}

 \title{Unresolved triples in the open cluster NGC~2437 (M~46) confirmed by the KMOS VVVX-GalCen Spectroscopic Survey}

 \subtitle{}

 \titlerunning{Discovery of a sequence of unresolved triples in NGC~2437}

 \author{D.~Minniti\inst{1,2}
 \thanks{Based on data from the ESO Public Survey program IDs
 179.B-2002 and 198.B-2004 taken with the VISTA telescope.},
 T.~Palma\inst{3}, 
 C.~O.~Obasi\inst{1}, 
  M.~G\'omez{\inst1}, 
 R.~K.~Saito\inst{4}, 
 J. G. Fern\'andez-Trincado\inst{5}, 
 J.~Alonso-Garc\'ia\inst{6}, 
 E.~R.~Garro\inst{7}, 
 J.~Corral-Santana\inst{7}, 
 A.~Luna\inst{7}, 
 P.~W.~Lucas\inst{8}, 
 M.~G.~Navarro\inst{9}, 
 V.~D.~Ivanov\inst{10}, 
 Z.~Guo\inst{11, 12}, 
 J.~Borissova\inst{11}, 
 R.~Kurtev\inst{11}, 
 C.~Morris\inst{11}, 
 V.~Fermiano\inst{11,12}, 
 J.~Osses\inst{11,12}, 
 P.~Esteves\inst{4}, 
 C.~C\'aceres\inst{1}, 
 B.~Dias\inst{1}, 
 S.~Saroon\inst{1}, 
 S.~Federle\inst{1}, 
 M.~Lad\inst{1}, 
 I.~Petralia\inst{1}, 
 P.~Rodriguez Cuevas\inst{1}, 
  J.~B.~Pullen\inst{1}, 
 D.~Bhadrakumar Sindhu\inst{6}, 
  L.~Gramajo\inst{3}, 
 L.~Baravalle\inst{3,13}, 
 D.~Galdeano\inst{14},
 M.~V.~Alonso\inst{3,13}, 
 \and
 J.~L.~Nilo Castellon\inst{15}
 }

 \authorrunning{Dante Minniti et al.}

 \institute{Instituto de Astrof\'isica, Depto. de. F\'isica y Astronom\'ia, Facultad de Ciencias Exactas, Universidad Andres Bello, Av. Fern\'andez Concha 700, Las Condes, Santiago, Chile 
\and Vatican Observatory, V00120 Vatican City State, Italy 
\and Observatorio Astronómico de Córdoba, Universidad Nacional de Córdoba, Laprida 854, X5000BGR, Córdoba, Argentina
\and Departamento de Física, Universidade Federal de Santa Catarina, Trindade 88040-900, Florianópolis, Brazil
\and Centro de Investigaci\'on en Astronomía, Facultad de Ingeniería, Ciencia y Tecnología, Universidad Bernardo O’Higgins, Avenida Viel 1497, Santiago, 8370993, Chile
\and Centro de Astronomía (CITEVA), Universidad de Antofagasta, Av. Angamos 601, Antofagasta, Chile
\and European Southern Observatory, Alonso de Córdova 3107, Vitacura, Santiago, Chile
\and Centre for Astrophysics Research, University of Hertfordshire, College Lane, Hatfield AL1 09A, UK
\and Istituto Nazionale di Astrofisica (INAF), Rome Astronomical Observatory, Via Frascati, 33, 00078 Monte Porzio Catone, RM, Italy
\and European Southern Observatory, Karl Schwarzschildstr 2, 85748 Garching bei München, Germany
\and Instituto de F\'isica y Astronom\'ia, Universidad de Valpara\'iso, Av. Gran Breta\~na 1111, Playa Ancha, Casilla 5030, Chile
\and Chinese Academy of Sciences South America Center for Astronomy (CASSACA), National Astronomical Observatories, CAS, Beijing 100101, China
\and Instituto de Astronom\'{\i}a Te\'orica y Experimental, (IATE-CONICET), Laprida 854, X5000BGR, C\'ordoba, Argentina
\and Departamento de Geofísica y Astronomía, CONICET, Facultad de Ciencias Exactas, Físicas y Naturales, Universidad Nacional de San Juan, Av. Ignacio de la Roza 590 (O), J5402DCS, Rivadavia, San Juan, Argentina
\and Departamento de Astronomía, Facultad de Ciencias, Universidad de La Serena. Av. Raul Bitran 1305, La Serena, Chile
}
\offprints{vvvdante@gmail.com}

\date{Received June xx, 2026; accepted XXXX xx, 2026}

 
 \abstract
 {  Recent optical and near-IR photometric databases like Gaia data release 3 (DR3) and the VISTA  Variables  in the  Vía  Láctea  eXtended (VVVX)  ESO  public survey allow the exploration of candidate unresolved multiple systems (binaries, triples, quadruples, etc.) in nearby clusters, that need spectroscopic confirmation. }
 { NGC~2437 (M~46) is a benchmark open cluster with well determined physical parameters. Its relative proximity ($D=1.6 \pm 0.1$ kpc) makes this cluster an ideal laboratory 
 to measure the fractions of unresolved multiple systems.   We aim here to  identify unresolved triple and
   higher  order systems  that are reliable members of the open cluster NGC~2437, and provide the spectroscopic confirmation using radial velocities.}
 {  The  selection  of cluster members  is made using  Gaia DR3 astrometry and optical photometry in combination with near-IR 
   photometry from the VVVX survey.  We  make optical  and  near-IR color-magnitude  diagrams to  select
   stars  located well  above  the  cluster   main-sequence,  which should include suitable unresolved multiple systems.  }
 { We  confirm  that there  is a  sequence of  unresolved binaries in  NGC~2437, and  report the discovery of a  sequence of unresolved  triple systems  in this
   cluster. We corroborate membership for 13 binary systems and 8  triple systems using radial velocities measured  by the  ESO KMOS  VVVX-GalCen
   spectroscopic  survey.   We further select stars with colors consistent with K-dwarfs, in order to have a controlled sample, comparable to the Solar  neighborhood. 
We also find that the fraction of unresolved triples and quadruples is  higher than  that of the older open cluster Trumpler~19, studied using a similar setup, suggesting that these systems merge to form blue stragglers or get dissolved with time.  }
 { We report the spectroscopic  confirmation of  95\% of the selected VVVX targets, thereby securing the presence of a  sequence of unresolved  triple systems in NGC~2437, located above the sequence of equal-mass binaries.
 The single stars constitute 60 \% of the mass of the cluster, while the rest of the mass (40 \%) is lodged in the unresolved multiple systems.
 }


 \maketitle
%

\section{Introduction}\label{Introduction}

There are  many different  channels  for the  formation and evolution  of hierarchical  systems with  comparable mass  components
\citep{Tokovinin2023}, and it is important  to increase the census of known triple systems  in open clusters  with a  wide variety of  ages. 
Finding large numbers of unresolved multiple systems in clusters would have implications  within  the simulations of the internal dynamical evolution for star clusters, as breaking loose  triples may help
to  cool down  the cluster,  and  also  favor  the   ejection  of  cluster  members.
Additionally, if triple  systems play a major  role in
the formation of  blue stragglers \citep{Fabrycky2007, Perets2009, Leigh2013}, the simple prediction  that can be made is that the number of  blue stragglers should be directly correlated with
the number of triple systems. 

Motivated by the success with the spectroscopic confirmation of the unresolved triple candidates in the open cluster Trumpler~19 \citep{Minniti2026}, we decided to look at other clusters with similar metallicity but closer to the Sun, in order to probe fainter magnitudes down the main sequence (MS).
Our main cluster selection criteria include to be located within the footprint of the VISTA  Variables  in the  Vía  Láctea  eXtended (VVVX)  ESO  public survey \citep{Saito2024}, older than 200 Myr,
closer than 2 kpc, not very reddened ($A_K<0.05$ mag), and containing enough astrometric members to acquire a sizable sample ($N>1500$ members). 
Very few clusters match these criteria, and we chose NGC~2437 (M~46) that is a benchmark cluster with known physical parameters:
$D=1.6 \pm 0.1$ kpc, age $t=355$ Myr, metallicity $[Fe/H]=-0.08$ \citep{Palma2026}.
In particular, it is enlightening to compare NGC~2437 with an older open cluster with similar  physical characteristics  like Trumpler~19  with age $t=3.7$ Gyr, and metallicity, $Z=-0.015$  
\citep{Dias2021, Sheikh2025, Minniti2026}. 
The study of NGC~2437 also enables the direct comparison with the sample of K-dwarfs in  the Solar neighborhood.
In order to better understand the mixed populations in the Solar neighborhood, astronomers have resorted to the study of star clusters, which are much simpler populations.
As opposed to the nearby K-dwarf sample that contains different ages, kinematics and chemical compositions \citet{Hubbard2026}, NGC~2437 offers a controlled sample where we can select stars with similar spectral types. This well-understood sample may serve as a reference to contrast with systematic multiplicity studies of the local population.  

NGC~2437 is located in the VVVX tile e0851, and covers about a square degree in the sky, and it is the most extended star cluster in the VVVX survey area. 
Most importantly, this open cluster has a sizable binary fraction ($\sim 30$ \%) according to \citet{Jadhav2021}, and \citet{Palma2026}.
In fact, during our previous work \citep{Palma2026}, we noticed that in the deep color-magnitude diagram (CMD) of this cluster there were many objects located above the sequence of equal-mass binaries that lies 0.75 mag on top of the MS.
These objects are prime candidates for being unresolved triple and higher order multiples, but this must be checked carefully because NGC~2437 is located at $GL= 231.89$ deg, $GB= 4.05$ deg and field contamination is important at low Galactic latitudes.

When searching for unresolved multiple systems, the turn-off region can be challenging to use because this is complicated not only by the fact that the isochrones become vertical, but also by the presence of blue stragglers, $\delta$Sct variables, multiple populations, and other sources of contamination.
The solution is to explore fainter stars down the MS, which is quite challenging in terms of telescope time. The KMOS VVVX-GalCen survey (Gómez et al., in prep.) allows us to reach fainter objects down the MS for this cluster.

This paper is organized as follows. 
Section 2 presents the photometric sample selection.
Section 3 describes the KMOS spectroscopy and radial velocity (RV) measurements. 
Section 4 discusses the results, including the comparisons with Trumpler 19 and the local K-dwarfs.
Finally, Section 5 presents the main conclusions.

\section{Photometric sample selection }

The selection for NGC~2437 members follows \citet{Palma2026} and \citet{Minniti2026}.
The only difference with the latter study is that we do not have photometry from the  Dark Energy  Camera Plane Survey 2  -- DECaPS2  \citep{Saydjari2023}.
Briefly, after matching the Gaia DR3 \citep{Drimmel2023} with the VVVX data, we used  photometry, proper motions (PMs) and parallaxes to discard non-members.

\begin{figure}[!]
\centering
\includegraphics[width=72mm]{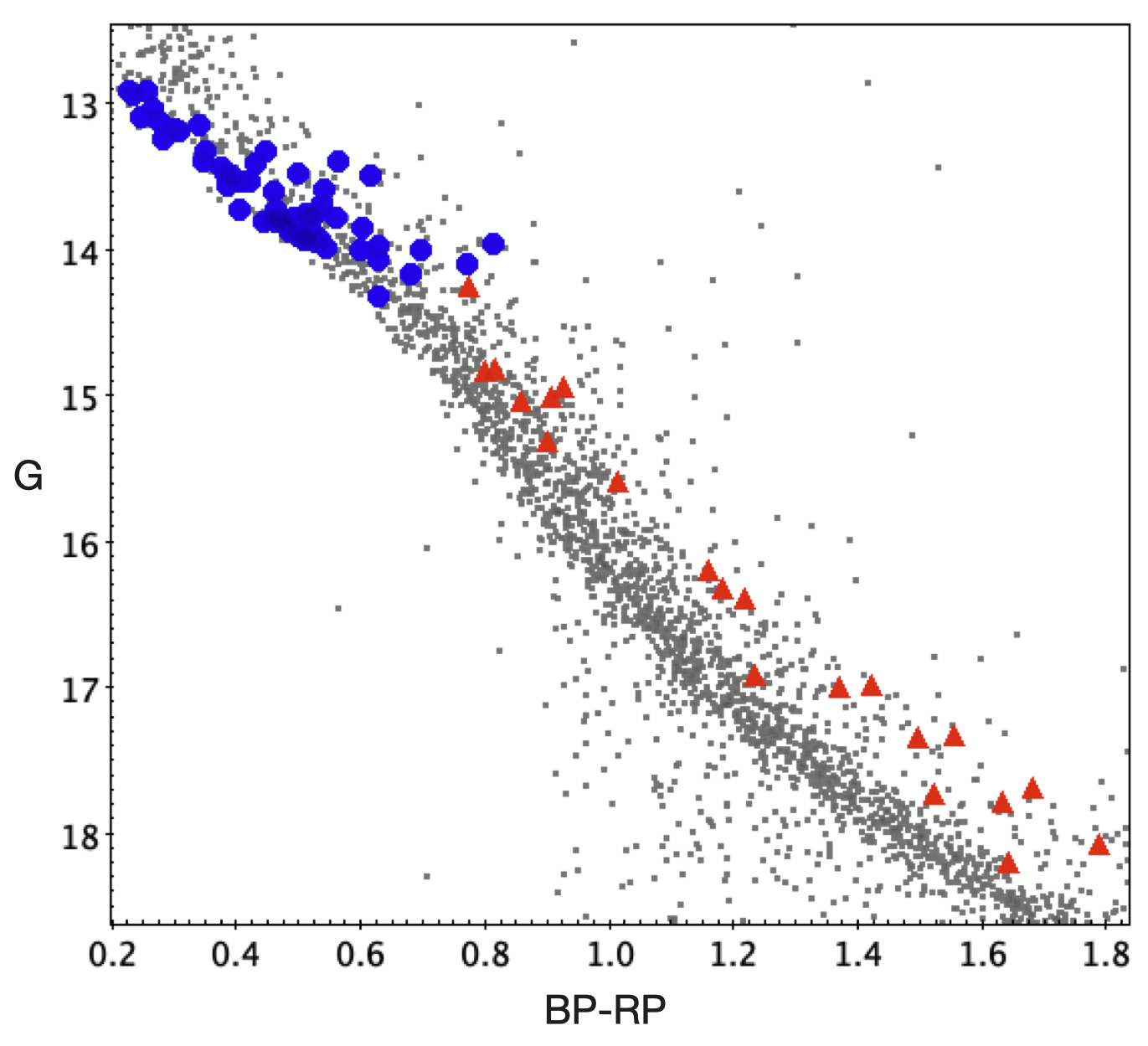}
\caption{ Gaia DR3 optical CMD for NGC~2437 members (selected from PMs and parallaxes) that have VVVX photometry. 
There are numerous objects above the cluster MS, that can be unresolved binaries, triples or quadruples. 
The RV targets observed with KMOS (red triangles) compared with the RV targets measured by Kiss et al. (2008, blue circles).  }
\label{fig:figura2}
\end{figure}

Unresolved multiple systems (binaries, triples, quadruples, etc.) are brighter than single stars
and are thus located above the main-sequence in the CMDs (Figure~\ref{fig:figura2}). Their total luminosity depends on the mass ratio q, which reaches a maximum of 1.0 in the case of equal-mass binaries. 
Therefore, the sequence of unresolved equal-mass binaries should
be located 0.75 mag above than the MS. Similarly, the unresolved
equal-mass triplets should be 1.19 mag brighter than the MS, the unresolved equal-mass quadruples should be 1.51 mag brighter than the MS, 
and the unresolved equal-mass quintuples should be 1.75 mag brighter than the MS, and so forth in the case of higher-
order multiple systems. As the likelihood of higher order multiple system diminishes, for simplicity we call triples to systems that contain three or more components.

Therefore, while searching for triples in NGC~2437 we consider that the triple systems should be brighter than the MS by $\Delta G, \Delta Ks >0.85$ mag, accounting for its intrinsic photometric width.
Note that the optical and near-IR photometric errors are quite small in comparison ($\sigma < 0.02$ mag -- Palma et al. 2026).
We are able to separate the cluster sequences taking advantage of the combination of the optical with the near-IR photometry. 
After fitting the MS in the optical and near-IR CMDs, we identified the location of the sequences of equal mass binaries, triples and quadruples.
Using the $Ks$ vs $BP-Ks$ as the most convenient reference CMD, we selected the candidate MS stars, binaries and higher order systems.

Contamination by field stars is a problem at low Galactic latitudes, as revealed by examining background fields \citep{Palma2026}. 
The foreground contamination is significant at around $J-Ks = 1.0$ mag, where the nearby M-dwarfs occupy this region of the CMD. 
The background contamination consists mostly of distant G-type dwarfs that concentrate at the intermediate-age turn-off around $J-Ks = 0.35$ mag.
In the intermediate color range $0.4<J-Ks<0.85$ mag, there is a sweet spot where the field contamination is minimized.
However, some interlopers may remain even after the stringent astrometric selection. 
In order to evaluate the residual contamination, we have selected two background fields located at the same latitude, but one degree away to the East and to the West of the cluster. 
After applying the same selection cuts in the parallaxes and PMs, we find that the average contamination is $<6\%$ in the relevant region of the CMDs.
Of course the "acid test" for contamination membership comes from the RVs, and as we will demonstrate in the next Section, this photometric contamination estimated using background fields agrees with the contamination fraction observed using the RVs.

\subsection{Variable stars in NGC2437}\label{Variables}

\begin{figure}[!]
\centering
\includegraphics[width=72mm]{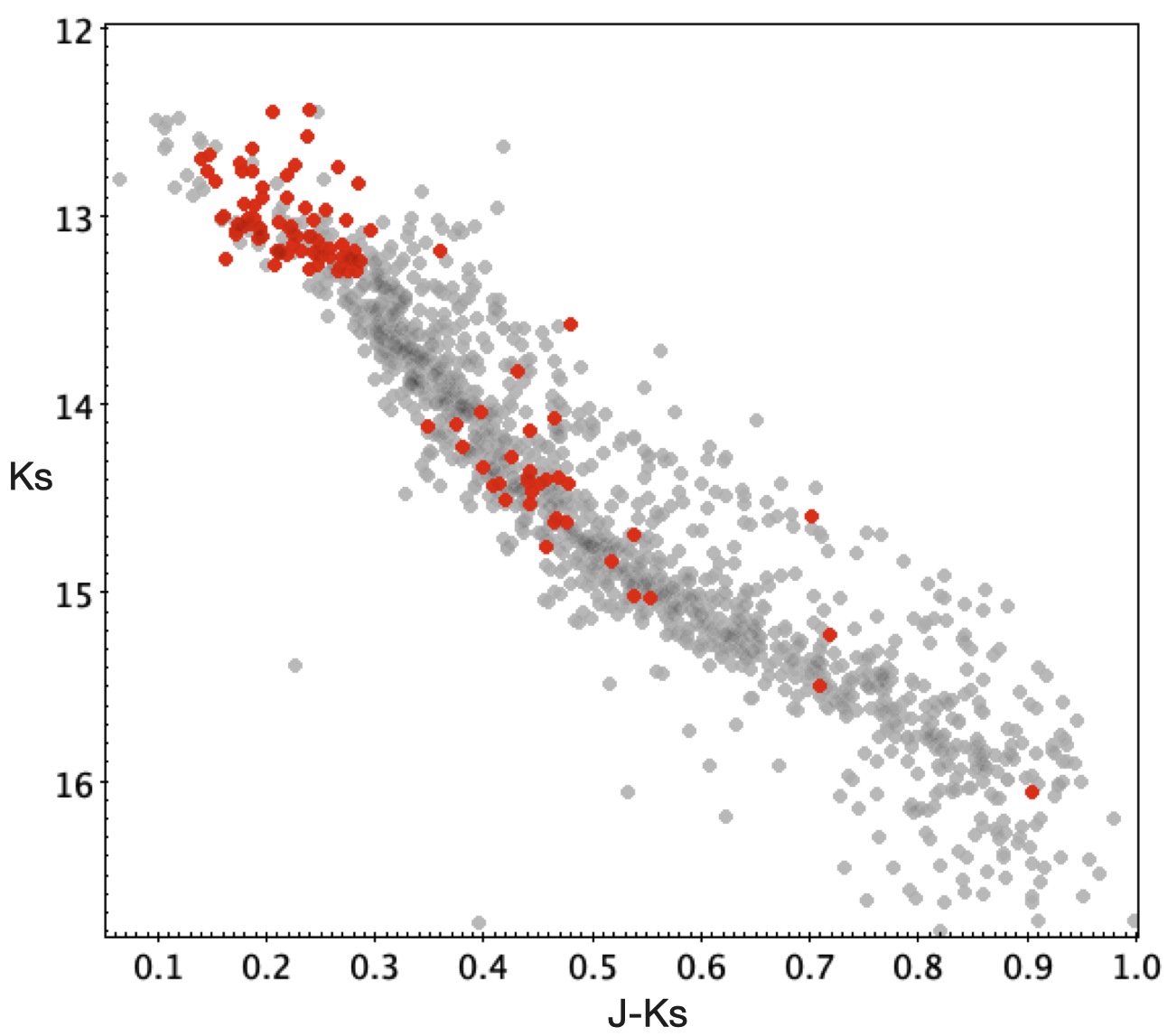}
\caption{Near-IR CMD showing Gaia DR3 selected members of NGC~2437 (grey circles).
The sources classified as variables in Gaia DR3  are highlighted in red. 
The concentration of variables in the upper part of this CMD corresponds  mostly to $\delta$ Sct oscillators, located within the instability strip.
There also is an excess of MS variables at $J-Ks \approx 0.44$ mag, $G-Ks \approx 1.8$ mag.
Aside from these regions,   very few variable sources are located well above the MS, in the region occupied by multiple systems. }
\label{fig:figura1}
\end{figure}

We also examined the population of  variable stars known in this cluster. 
There are $Nvar=36$ candidate variable sources from Gaia DR3 in total, including 
$Nvm=27$ variable MS stars (2.1\% of the total Nms),
$Nvb=7$ variable binaries (1.8\% of the total Nbin),
and
$Nvt=2$ variable triples (8\% of the total Ntri).
These variable sources are not distributed evenly in the cluster CMD (Figure~\ref{fig:figura1}). 
The upper part of the  CMD shows a clear concentration of variables within the instability strip, that should comprise mostly $\delta$ Sct oscillators. 
Note that there also appears to be an excess of MS variables at $J-Ks \approx 0.44$ mag, $G-Ks \approx 1.8$ mag, that deserve further examination.
We find that aside from these regions, there are very few variables located well above the MS, in the region occupied by unresolved binary and triple star candidates.
In other words, variability cannot explain the sources located above the MS, which should be unresolved multiple systems.

\section{The KMOS spectroscopy }

The  near-IR  spectroscopy  was performed  with  the  KMOS
spectrograph  mounted   at  the  Very Large Telescope (VLT) UT1  located  at   ESO  Paranal
Observatory,  as part  of the  ESO Large  Spectroscopic Public  Survey
VVVX-GalCen (G\'omez et  al.,  in  prep.). 
  We selected   22 objects located well above the MS (unresolved binary and triple candidates) for  spectroscopy. 
Figure~\ref{fig:figura2} shows  the selected  stars for KMOS spectroscopy in red, compared with the stars observed spectroscopically by \citet{Kiss2008} in blue.
The latter occupy the upper portion of the MS, both datasets are very complementary, with the main difference (aside from the obvious magnitude difference) being that the spatial coverage  of our KMOS observations 
is more concentrated than that of  \citet{Kiss2008}.
These RVs can also be complemented with the Gaia RVs measured  for bright stars ($G<15$ mag) in the cluster region. 

\begin{figure*}[!]
\centering
\includegraphics[width=132mm]{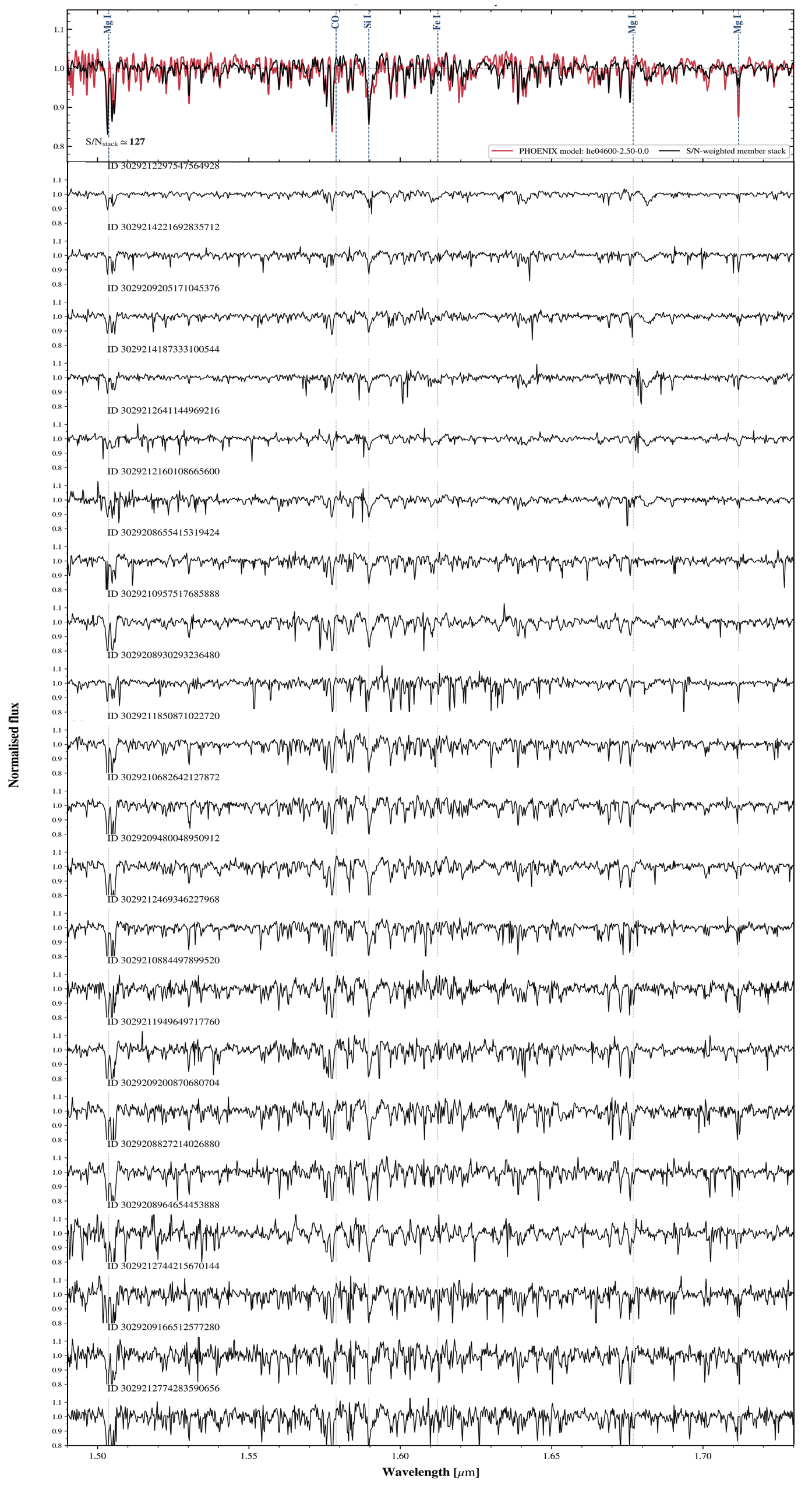}
\caption{ 
H-band KMOS spectra (R= 4,000) for the candidate unresolved multiple systems in NGC2437 (see Table 1). 
The top panel highlights the stacked weighted spectrum (total S /N= 61) compared with a PHOENIX LTE model from \citet{Husser2013} for $T_{eff} = 4300$ K, $log ~g= 4.0$ and solar metallicity (red line).
Some of the prominent spectral features are labelled,  
The bottom panel shows the normalized individual spectra, evenly displaced vertically. Increased scatter can be observed in regions where telluric features are present. 
}
\label{fig:figura3}
\end{figure*}

The
selection among  the candidate targets  was made in order  to optimize
the  KMOS  arm  allocation,  and  the  targets  with  measured  RVs are  located well within the  core of the cluster  to fit
the  instrumental  field  of  viewof KMOS  (approximately $7.5$ sq.arcmin.).   The  near-IR
spectra were acquired with  a single  KMOS pointing  in the  $H$-band configuration,
using a total exposure time of 58~min  starting at UT~23:01:30 on the night
of 17  May 2026, under $0.6^{\prime\prime}$~seeing conditions.   The spectra were
reduced and calibrated  using the  ESO Data Processing System  (EDPS; Freudling et al. 2004) with the dedicated KMOS pipeline, resulting in
22  good spectra with  $15 < S/N < 50$. 
Figure~\ref{fig:figura3} illustrates the normalized spectra, highlighting some strong features.

The RVs were  measured using
cross correlations \citep{Tonry1979},  comparing with a PHOENIX LTE model from \citet{Husser2013} for $T_{eff} = 4300$ K, $log g= 4.0$ and solar metallicity.
We note that  due to the moderate  resolution of KMOS  ($R = 4,000$), it would be difficult
to discriminate  most spectroscopic  binaries or to  measure rotations
from   line  profiles,. However, it  is possible  to measure reliable RVs,
accurate  to  $\sigma_{RV}<4.5$ km~s$^{-1}$,  that are useful to  confirm  cluster
membership.   

\begin{figure}[!]
\centering
\includegraphics[width=82mm]{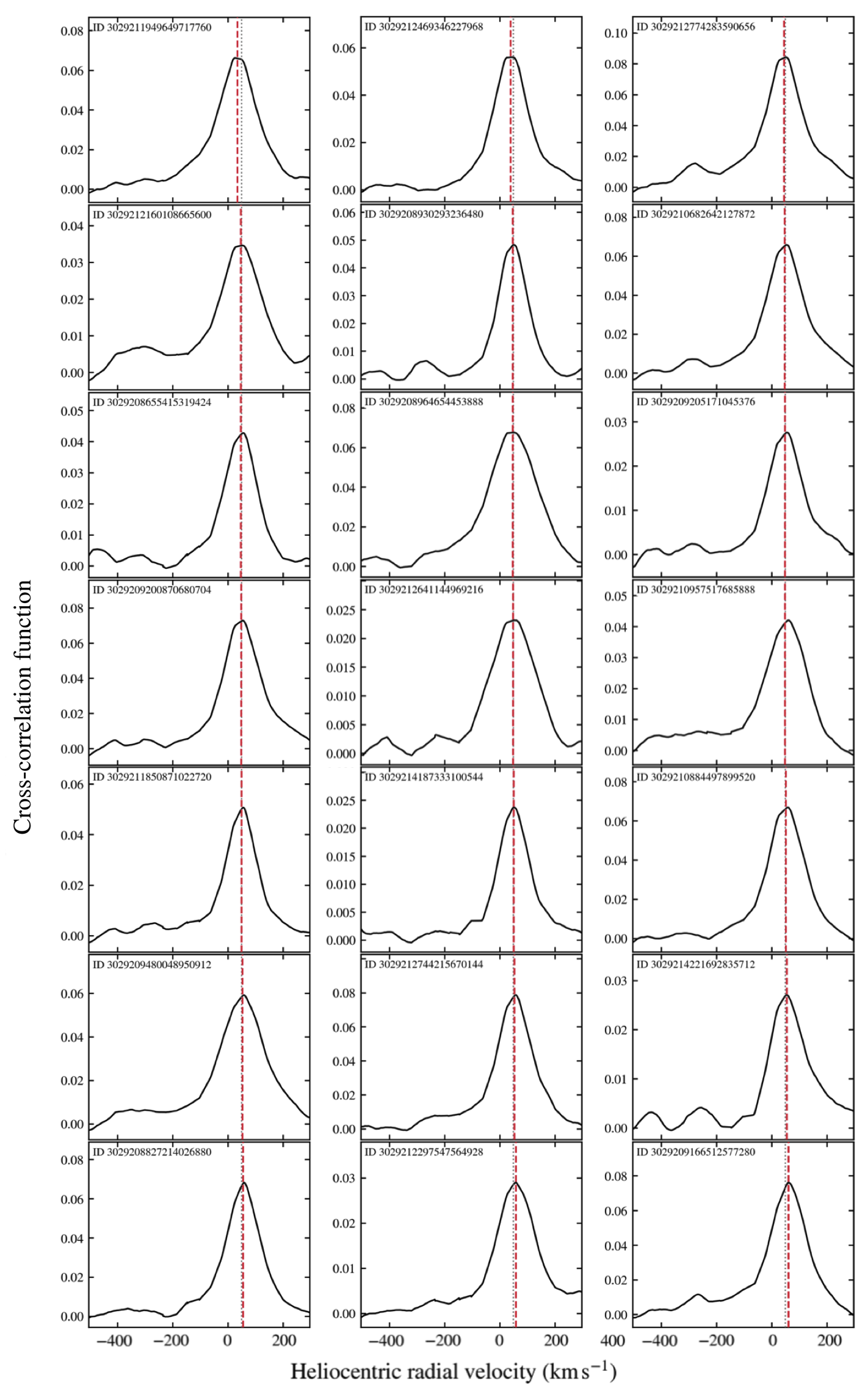}
\caption{ 
Peaks of the CCFs measured from the KMOS spec-
tra of NGC~2437 members listed in Table 1. The vertical
lines indicate the measured RV (dashed red line)
compared with the mean cluster velocity (dotted black line).
}
\label{fig:figura5}
\end{figure}

The peaks of the cross correlation functions (CCFs) for the NGC~2437 member stars are shown in Figure~\ref{fig:figura5} .
These CCF peaks are well defined and no double peaks were detected. 
At the KMOS resolution, we are not able to reliably fit multiple components, as done for example by   \citet{Merle2017} and  \citet{Kounkel2021} 
for their high resolution  CCFs.
In the absence of significant rotation, empirically we determine that a single velocity star (or a multiple system seen pole-on) should produce a narrow peak, with measured FWHM ~120 km~s$^{-1}$. 
However, it is evident that most of the CCF peaks are significantly broader than this. 
We count 5 stars with narrow CCFs, and 16 stars with broader peaks, indicative of high rotation or multiple velocities. 
This is consistent with the argument based on the positions in the CMDs that the majority (if not all) of our targets are binaries or triples. 
In fact, even the narrow peak objects may be multiple, as the measured fraction is consistent with the expected proportion of random orbits that would be observed close to pole-on.

\begin{figure*}[!]
\centering
\includegraphics[width=162mm]{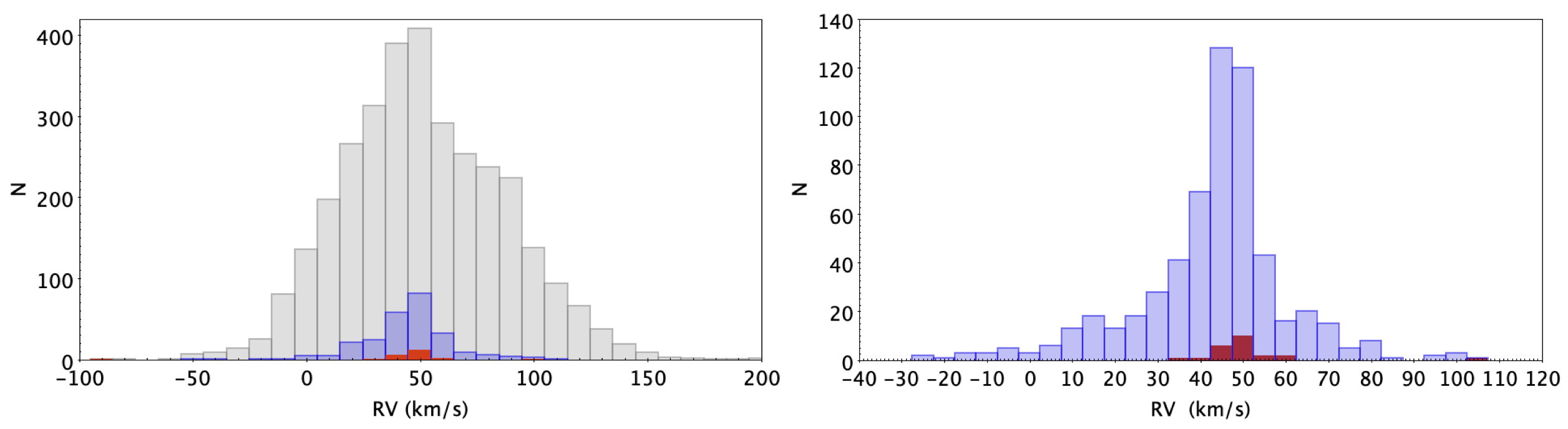}
\caption{ Left panel: KMOS RV distribution (red) compared with the Gaia RVs (grey = all stars, blue = PM cluster members). Right panel: KMOS RV distribution (red) compared with the AAOMEGA RVs from Kiss et al. 2008 (blue).  }
\label{fig:figura4}
\end{figure*}

Figure~\ref{fig:figura4}   shows  the   measured  RV distribution, which  exhibits a small dispersion  ($\sigma = 5.6
\pm 1.2$~km~s$^{-1}$) that is similar to other clusters observed at similar resolution and S/N for our
VVVX-GalCen spectroscopic survey, from which we conclude  that all but one star (Gaia DR3 3029212778583872896) are
confirmed members of the cluster. We  measure a mean velocity for this
cluster $<RV>= 48.6  \pm 2.0$~km~s$^{-1}$, in good  agreement with the
Gaia    mean   velocity    $<RV>   =    48.1   \pm    0.5$~km~s$^{-1}$
\citep{Drimmel2023}.   Table~1  lists  the photometry and  measured RVs  for
the NGC~2437 targets  observed,  including the   star  that is
considered non-member (Gaia DR3 3029212778583872896).  
Particularly noteworthy is the  case of the
$K_{\rm s}=  12.88$~mag star Gaia DR3  3029212297547564928, located at
$RA=115.452051$~deg,  $Dec= -14.807121$~deg, that is  variable in the
Gaia DR3 catalog.   This target has a large RUWE=1.87 but
complies with  all of the selection  cuts, and it can  be considered a
cluster  member  according   to  the  $RV=57.8  \pm 1.35$~km~s$^{-1}$
(spectrum $S/N=48.3$).

Since these are among the first RVs measured in the near-IR by the KMOS VVVX-GalCen survey, it is important to make external comparisons with previous works for NGC~2437.
For this purpose we chose to compare our KMOS RVs with the optical RVs measured by Gaia DR3 (Gaia Collaboration 2023), 
by the AAOMEGA spectrograph \citep{Kiss2008}, and by the Hydra spectrograph \citep{Frinchaboy2008}.
Note that our spectra have lower resolution $R_{KMOS}=4,000$ than the spectra from  \citet{Kiss2008}  with $R_{AAOMEGA}=8,000$, and than the spectra from \citep{Frinchaboy2008} with $R_{Hydra}=15,000$.
 \citet{Kiss2008}  report RVs accurate within $\pm 1-2$ km~s$^{-1}$ for the cooler stars and $\pm 5$ km~s$^{-1}$ for the hotter stars in their sample, while \citet{Frinchaboy2008} report even smaller RV errors. 

The RVs for the sample cluster members  listed in Table 1 range between 32 km~s$^{-1}$ and 57 km~s$^{-1}$.
Note that discrepant velocities do not necessarily exclude cluster membership, because they could be the result of orbital motions. 
However, in the absence of additional evidence, stars outside this RV range are not considered.
Our mean velocity $<RV>=48.6 \pm 2.0$ km~s$^{-1}$, and  velocity dispersion $\sigma = 5.5 \pm 1.2$ km~s$^{-1}$,
are in excellent agreement with those of  \citet{Kiss2008}, that obtained
$<RV>=48$ km~s$^{-1}$, and $\sigma = 3.9 \pm 0.3$ km~s$^{-1}$, respectively (using $N \approx 300$ cluster members). 
They pointed out that the relatively large velocity dispersion for an open cluster suggests a significant fraction of binaries.
These values agree in turn with the high dispersion spectroscopy for 19 stars from \citet{Frinchaboy2008} that obtained
$<RV>=46.9 \pm 1.0$ km~s$^{-1}$.

Additionally our targets are  unresolved multiple star candidates, so the velocity dispersion may not be representative of the whole cluster population if there has been significant dynamical evolution.
When comparing with our own KMOS RVs for the open cluster Trumpler~19, we find that the velocity dispersion of the two clusters are slightly different: $\sigma_{NGC2437} < \sigma_{Trumpler19}$, although this can be due to the higher $S/N$, and therefore better RVs, measured for NGC~2437.

This external comparison yields another interesting result, because  we found one source from from  \citet{Kiss2008}  that we define as a likely triple candidate.
This is Gaia DR3 3029161307695619456, located at Galactic coordinates $GL= 232.06747$ deg, $GB= 4.23019 $ deg. The Gaia DR3 astrometry for this source yields
$Parallax=0.768 \pm 0.086$ mas and $Parallax/error=9.0$, implying a distance of $D=1302$ pc,  
$PMRA=-3.85 \pm 0.08$ mas/yr, and $PMDEC=0.60 \pm 0.08$ mas/yr.
 \citet{Kiss2008} obtain  $RV_{AAOMEGA}= 55.3  \pm 5$ km~s$^{-1}$, that secures the cluster membership for this object.
Also note that this star is not flagged as a variable by Gaia DR3.
Inspecting the available images, this source appears as a perfect point source in PANSTARRS and VVVX images, but the very high re-normalized unit weight error $RUWE= 4.94$ is also consistent with an unresolved multiple system.
The photometric parameters for this object are
$G=13.96$ mag,  $BP-RP=0.81$ mag from Gaia DR3, and  $J= 12.87$ mag,  $H= 12.66$ mag,   $Ks=12.54$ mag,   ($+/-0.01$ mag) from the VVVX survey (relative source separation $= 0.06$"), yielding $G-Ks= 1.42$ mag, and $BP-Ks= 1.73$ mag. 
This star is shown with a large circle in the CMD of Figure~\ref{fig:figura6}.
The physical parameters listed in the Gaia DR3 database are $T_{eff}= 7067$ K,  $log ~g=4.29$,  and $[Fe/H]= -0.42$. 
This Gaia metallicity is low compared with the adopted mean cluster  metallicity $[Fe/H]=-0.08$ \citep{Palma2026}, and the derived reddening  $E(B-V)= 0.38$ mag is somewhat higher than the cluster mean reddening.
Nonetheless, we conclude that Gaia DR3 3029161307695619456 is a bonafide unresolved triple system in NGC~2437.

%
\begin{table*}
 \caption{NGC~2437 candidate unresolved multiple systems.
}
\label{tab:data}
\centering
\scriptsize
\begin{tabular}{l c c c c c c c c c c c c c}
\hline
\hline
\textbf{Gaia DR3 ID} & 
\textbf{RA} & 
\textbf{DEC} & 
\textbf{$G$} & 
\textbf{$BP$}  & 
\textbf{$RP$}  & 
\textbf{$J$} & 
\textbf{$K_{\rm s}$}&
\textbf{$RV$}&
\textbf{Comments}\\
&
 \textbf{(deg)} &
 \textbf{(deg)} &
 \textbf{(mag)} &
 \textbf{(mag)} &
 \textbf{(mag)} &
 \textbf{(mag)} &
 \textbf{(mag)} &
 \textbf{(km s$^{-1}$)}
  \textbf{} &\\
\hline
3029208655415319424  &  115.46712754726288  &  -14.859592390761511 &15.620  &  16.037  &   15.021  &  14.294  &  13.846 & 46.5  $\pm$ 1.4  &  binary member  \\   
3029208827214026880  &  115.45219591494187  &  -14.855977692183714 &17.031  &  17.655  &   16.285  &  15.287  &  14.634 & 55.0  $\pm$ 1.1  &  binary member    \\
3029208930293236480  &  115.44561013504833  &  -14.838093766619203 &15.062  &  15.371  &   14.513  &  13.994  &  13.609 & 46.4  $\pm$ 0.5  &  binary member     \\
3029208964654453888  &  115.48739833364729  &  -14.861983174761859 &17.365  &  18.065  &   16.569  &  15.523  &  14.807 & 46.7  $\pm$ 2.0  &  binary member      \\
3029209166512577280  &  115.50033929878231  &  -14.830513531939946 &18.224  &  19.029  &   17.385  &  16.173  &  15.393 & 60.4  $\pm$ 2.0  &  binary member    \\
3029209200870680704  &  115.50885755194790  &  -14.809644186757131  &17.714  &  18.526  &   16.843  &  15.659  &  14.845 & 47.7  $\pm$ 1.2  &  binary member    \\
3029209205171045376  &  115.50938663459470  &  -14.82019208362855    &14.849  &  15.171  &   14.355  &  13.774  &  13.401 & 47.5  $\pm$ 0.8  &  triple member      \\
3029209480048950912  &  115.49113215100158  &  -14.788821921142054 &17.009  &  17.637  &   16.216  &  15.142   &  14.430 & 52.2  $\pm$ 2.0  &  triple member     \\
3029210682642127872  &  115.51619356809131  &  -14.803658474262456 &16.941  &  17.081  &   15.844  &  14.996  &  14.390 & 46.4  $\pm$ 1.1  &  triple member     \\
3029210884497899520  &  115.51216812087895  &  -14.79535132658909   &17.820  &  18.632  &   16.998  &  15.751  &  14.929 & 51.2  $\pm$ 2.0  &  binary member     \\
3029210957517685888  &  115.50195953133466  &  -14.787276764636955 &16.354  &  16.861  &   15.679  &  14.830  &  14.270 & 47.9  $\pm$ 10.0  &  triple -member\\ 
3029211850871022720  &  115.42713342530458  &  -14.838476611566161 &16.422  &  16.956  &   15.737   &  14.864  &  14.316 & 48.6  $\pm$ 1.2  &  triple member    \\
3029211949649717760  &  115.42465855267181  &  -14.821144562602584 &17.359  &  18.096  &   16.539   &  15.427  &  14.687 & 34.7  $\pm$ 1.3  &  triple member      \\
3029212160108665600  &  115.40990836833743  &  -14.807457341400307 &14.975  &  15.348  &   14.421  &  13.760  &  13.367 & 46.2  $\pm$ 1.7  &  triple member     \\
3029212297547564928  &  115.45205145560249  &  -14.807121371487423 &14.283  &  14.583  &   13.809  &  13.248  &  12.881 & 57.8  $\pm$ 1.3  &triple member variable\\     
3029212469346227968  &  115.45980161259891  &  -14.783690648992065 &16.230  &  16.735  &   15.575  &  14.705  &  14.159 & 38.6  $\pm$ 1.1  &  binary member     \\
3029212641144969216  &  115.41729030355735  &  -14.778956195936678 &15.037  &  15.405  &   14.499  &  13.850  &  13.488 & 47.8  $\pm$ 1.9  &  binary member    \\
3029212744215670144  &  115.43796827567017  &  -14.7644305521707     &17.754  &  18.448  &   16.926  &  15.802  &  15.101 & 52.3  $\pm$ 0.7  &  binary member     \\
3029212774283590656  &  115.45080045424628  &  -14.753691430215401 &18.102  &  18.959  &   17.171  &  15.954  &  15.133 & 44.0  $\pm$ 1.3  &  binary member     \\
3029212778583872896  &  115.44175332893161  &  -14.754294368905263 &16.209  &   --          &   --          &  14.667  &  14.179 & 103.3 $\pm$ 1.0  &  non-member \\
3029214187333100544  &  115.47536786172434  &  -14.755720752085915 &14.852  &  15.169  &   14.367  &  13.800  &  13.517 & 49.4  $\pm$ 1.3  &  binary member    \\ 
3029214221692835712  &  115.47326464438767  &  -14.747675314700492 &15.342  &  15.704  &   14.804  &  14.145  &  13.789 & 53.7  $\pm$ 1.2  &  binary member    \\
\hline
\end{tabular}
\tablebib{The  target highlighted in bold is classified as a variable star in the Gaia DR3 database.}
\end{table*}
%

\section{Results and Discussion}

\begin{figure*}[!]
\centering
\includegraphics[width=162mm]{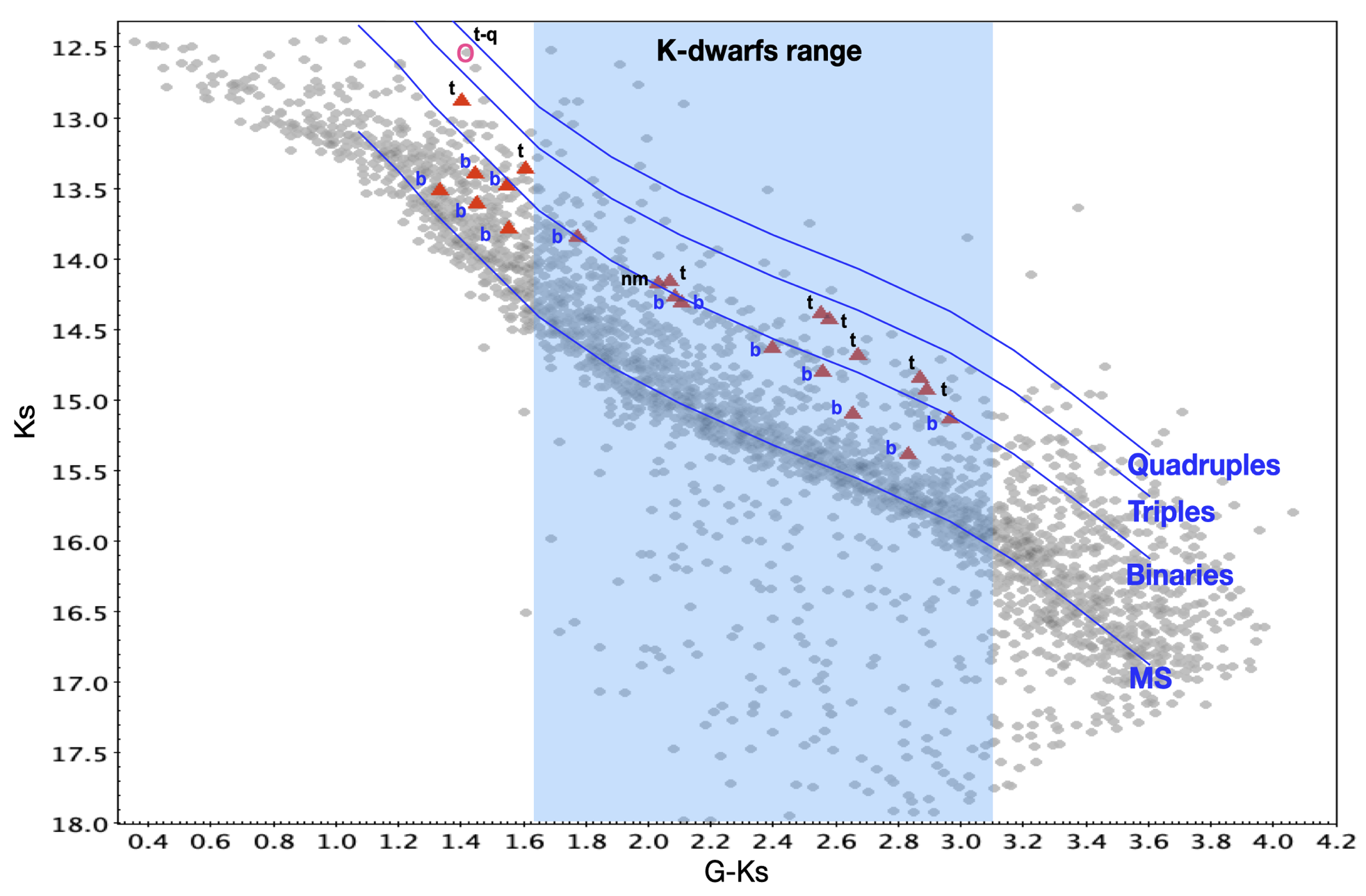}
\caption{ CMD summarizing the results from the KMOS RV targets in NGC~2437 (red triangles), confirming 8 triple candidates, 13 binaries, plus one non-member.
In addition, the circled star on top is a triple system that we have identified in the sample of  \citet{Kiss2008}.
The expected locations of the sequences of unresolved equal-mass multiple systems are labeled.}
\label{fig:figura6}
\end{figure*}

The  $Ks$ vs $G-Ks$ CMD of Figure~\ref{fig:figura6} summarizes the KMOS results, and Table 1 lists our final classification for each spectroscopic target, including 13 binary members, 8 triple members (plus one more from Kiss et al. 2008), and one non-member according to the RVs.
The results then include the  discovery of the sequence of unresolved triples  in NGC~2437, and the
spectroscopic confirmation  through RVs that the majority ($>95$\%) of the binary and triple candidates are cluster members.
Therefore, we can confidently now estimate the relative numbers of K-dwarfs that belong to the main, binary, and triple sequences. 
For a total sample of $N=1673$ members  selected using Gaia DR3 PMs and parallaxes  \citep{Palma2026}, we find 
$Nms=1258$ candidate MS objects, 
$Nbin=390$ candidate binary systems,
and
$Ntri=25$ candidate triple and higher order systems.

Using the definitions of \citet{Offner2023}, these yield a multiple fraction $MF=0.248$ (25\%), including
a binary fraction $BF=0.233$ (23\%), and
a fraction of triple and higher order systems  $THF=0.015$ (1.5\%).
These values are consistent with previous estimates of the binary fraction that did not consider higher order multiple systems \citep{Jadhav2021, Palma2026}.
There is some residual field star contamination, mostly from red dwarfs that are much more distant than the cluster, as shown by the presence of sources well below the MS in Figure 6. Some of these sources would be better discriminated using the improved results from the upcoming Gaia DR4 release. In addition, we expect that the finer astrometry may help to confirm the multiplicity detection for some of the bright sources as well.

 \citet{Kiss2008}  measured the cluster mass $M_{dyn}=110,000 ~M_{\odot}$, based on their RVs, and they argue that this is an overestimation because the total luminosity is much smaller $L = 8800 ~L_{\odot}$, indicating that this inflated dynamical mass is due to a large population of binary systems. Regardless of the exact total cluster mass,
the multiple fractions found here imply that single stars constitute 60\% of the mass of the cluster, while the rest of the mass (40\%) is lodged in multiple systems (binaries, triples, quadruples, etc.).  
NGC~2437 appears to be one of the clusters with best known parameters, that can be used as a benchmark cluster in future comparisons with the field and other open clusters.

\subsection{Comparison with Trumpler~19}
We have previously discovered a sequence of unresolved triples in the aged open cluster Trumpler~19  \citep{Minniti2026}.
NGC~2437 is much younger ($t= 355$ Myr) in comparison with Trumpler~19 ($t= 3.7$ Gyr), and also
much closer $D= 1.6$ kpc $vs$ $D= 3.5$ kpc, and less reddened $Av= 0.3$ mag $vs$ $Av= 0.77$ mag, respectively.
Thus, in practice NGC~2437 is less observationally challenging than Trumpler~19. Due to the distance and reddening difference, the observed relative magnitude offset between two stars of similar masses members of these clusters is more than 2 magnitudes (i.e., using the same setup, we can reach more than 2 magnitudes deeper down the MS of NGC~2437 than that of Trumpler~19).

Therefore, it is a great advantage that the VVVX near-IR photometry can reach fainter objects in NGC~2437, down in the MS to about $0.2 R_{\odot}$, $0.2 M_{\odot}$.
Unfortunately, NGC~2437 is outside the footprint of the  DECAPS2 survey   \citep{Saydjari2023}, so there is no deep optical photometry available.
Regarding the structural parameters, these clusters have similar sizes ($R_c= 2.74$ pc), and comparable masses,
determining that NGC~2437 is a slightly denser cluster.

\subsection{Comparison with local K-dwarfs}
It is  interesting to compare with the Solar neighborhood, where the advantage is that these multiple systems can be directly resolved. 
However, the nearby population is very heterogeneous in age and chemical composition, and 
NGC~2437 offers a controlled environment (volume limited sample with a single age and composition), that can in turn help us understand the multiplicity architecture observed in the Solar vicinity.
\citet{Hubbard2026} presented the spectroscopic results of a volume limited sample ($R<33.3$ pc) of 580 K-dwarfs. They derived the stellar properties, activity, and ages calibrated with 35 benchmark K-dwarfs with ages ranging from 20 Myr to 5 Gyr. Based on the H${\alpha}$ emission, Lithium absorption, and/or membership to young moving groups, they found that 8.8\% of the stars have signatures of youth, and also that 4\% of the stars appear to be metal-poor with $[Fe/H] < -0.50$. Exploring the kinematics, they conclude that the vast majority of the stars belong to the disk of the Galaxy (80\% thin disk, 18.4\% thick disk), with only one halo star in their sample. 
They discuss many cases of spectroscopically confirmed binaries/triples, but unfortunately do not include the statistics for the multiplicity status of their whole sample.
On the other hand, \citet{Shariat2025} present a large catalog of resolved triple systems within 100 pc of the Sun based on Gaia observations. They find a strong dependence of the intrinsic triple fraction with mass, increasing for larger masses. They obtain an  intrinsic triple fraction of ~$5 \%$ at low masses ($0.5 M_{\odot}$), which is in reasonable agreement with the present value for NGC~2437.

\section{Conclusions}

After carefully selecting the members based on the astrometry provided by the Gaia DR3, the different optical and near-IR CMDs for NGC~2437 (Figures 1 and 6) 
clearly show that there is a rich population of unresolved multiple systems (binaries, triples and quadruples) above the MS of this cluster. 
In order to verify the cluster membership of these candidate multiple systems, we have measured RVs from the VVVX-GalCen survey with KMOS at the VLT.

The results from the spectroscopy are that 
$N=22$ candidate multiple systems (binaries and triples) were observed with high S/N ($15<S/N<50$). 
These yield individual RVs good to $ \sigma_{RV} < 4.5$ km~s$^{-1}$. 
In total, $21$ are found to be confirmed RV members, including $13$ binaries, and 
$8$ triples. In addition there is another triple source with membership confirmed with RVs from the AAOMEGA spectroscopic sample of  \citet{Kiss2008}.
Seeing that only one source out of 22 is found to be a non-member according to the RVs, we conclude that the overall sample contamination is $\leq 5\%$.

Based on these results, we estimate the relative numbers of K-dwarfs that belong to the different sequences (MS, binaries, triples and quadruples).
We conclude that  single stars make up about 60\% of the mass of the cluster, while the rest of
the mass is located in unresolved multiple systems.
Comparing with the older cluster Trumpler~19 from our similar study, we find that NGC~2437 contains a higher proportion of multiple systems.
While in comparison with the Solar neighborhood, we find that NGC~2437  contains a similar proportion of multiple systems.
This work shows the path to study the multiplicity in other open clusters with known ages, kinematics and chemical compositions, in order to understand the origin of the population observed in the vicinity of the Sun.

Also, if triple  systems play an important  role in the formation of  blue stragglers, the implication is that the number of  blue stragglers should be directly correlated with the number of triple systems.
In  the case of NGC~2437, there are large numbers of both blue straggler stars and triple systems, as was recently observed in the older open cluster Trumpler~19 (Minniti et al. 2026).
While these results are encouraging, the number of open clusters with confirmed triple systems is very scarce, and more spectroscopic follow-up observations are needed.

Finally,  having identified a population of unresolved multiple systems in NGC~2437, it is important to follow them up with high-resolution multi-epoch spectroscopy to measure accurate RVs, and find the systems where orbital solutions can be determined in order to measure the masses and unveil their architecture. 
Deeper photometric observations would also be useful to extend this work into the M-dwarf regime, in order to determine the mass dependence of the fractions of unresolved binaries and multiples for NGC~2437 and other benchmark open clusters. 

\begin{acknowledgements}

We gratefully acknowledge  the use of data from the  ESO Public Survey
program IDs 179.B-2002  and 198.B-2004 taken with  the VISTA telescope
and data  products from the Cambridge Astronomical Survey  Unit. 
This research has made use of the SIMBAD database, CDS, Strasbourg
Astronomical Observatory, France, and of data from the European Space Agency (ESA) mission \textit{Gaia}
(\url{http://www.cosmos.esa.int/gaia}), processed by the \textit{Gaia}
Data Processing and Analysis Consortium (DPAC, \url{http://www.cosmos.esa.int/web/gaia/dpac/consortium}).
 C.C.,  B.D., and D.M. are supported by the ANID BASAL Center for  Astrophysics and Associated Tecnologies (CATA) project FB210003.  
T.P. is supported by Secretaría de Ciencia y Tecnología de la Universidad Nacional de Córdoba (SECYT-UNC, Res. 258/53).
 R.K.S. is supported by CNPq/Brazil   through  projects 308298/2022-5 and 421034/2023-8.
 C.O.O. is supported by the 2025 Postdoctoral Talent Attraction Competition for Research Centres and Institutes of Universidad Andres Bello, Project No. DI-07-25/ATP.
 M.G.  and S.F. are  supported by ANID Fondecyt  Regular No. 1240755.  
 J.A.-G.  is supported by  ANID Fondecyt Regular  No.   1201490.  
 J.G.F-T  is supported by ANID Fondecyt Regular No. 1260371, by the Joint Committee ESO-Government of Chile under the agreement 2023 ORP 062/2023, and by the Doctoral Program in Artificial Intelligence, DISC-UCN.
 P.W.L. is supported by STFC grant ST/Y000846/1.  
 R.K. and J.B. are supported by ANID Fondecyt Regular grants No. 1261142 and 1240249.
 E.R.G. and A.L.  gratefully  acknowledge   the  ESO Fellowship program.  
 Z.G. is supported by ANID Fondecyt Iniciacion  No. 11260176.  
 We  also acknowledge support by the China-Chile Joint Research Fund (CCJRF No.2301) and the Chinese  Academy  of  Sciences  South  America  Center  for  Astronomy (CASSACA) Key Research Project E52H540301.
 T.P. is supported by the Argentinian  Secretaría de  Ciencia  y  Tecnología de  la Universidad  Nacional de  Córdoba (SECYT-UNC,  Res. 258/53).  
I.P. is supported by ANID Becas Doctorado Nacional N. 21230761.
 J.O. is supported by ANID Becas/Doctorado nacional No. 2026-21262687.

\end{acknowledgements}

\makeatletter
\renewcommand\NAT@nmfmt[1]{\textsc{#1}}
\def\maxbibnames{2}  
\makeatother

\bibliographystyle{aa}

\end{document}